\documentclass[%
 reprint,
superscriptaddress,
 amsmath,amssymb,
 aps,floatfix
]{revtex4-2}

\usepackage{comment}
\usepackage{graphicx}
\usepackage{dcolumn}
\usepackage{bm}

\usepackage{xcolor}
\usepackage{siunitx}
\usepackage{soul}

\newcommand{\br}[1]{\left( #1\right)}

\renewcommand{\tensor}[1]{\underline{\underline{#1}}}
\renewcommand{\vec}[1]{\boldsymbol{#1}}
\newcommand{\uvec}[1]{\vec{\hat{#1}}}
\newcommand{\sAC}{\mathcal{C}}
\newcommand{\mean}[1]{\left\langle #1 \right\rangle}
\newcommand{\abgd}{\alpha \beta \gamma \delta }

\begin{document}

\preprint{APS/123-QED}

\title{Long ranged stress correlations in the hard sphere liquid}
\date{\today}
\author{Niklas Grimm}
\affiliation{Fachbereich Physik, Universit\"at Konstanz, 78457 Konstanz, Germany}
\author{Martin von Bischopinck}
\affiliation{Fachbereich Chemie, Universit\"at Konstanz, 78457 Konstanz, Germany}
\author{Andreas Zumbusch}
\affiliation{Fachbereich Chemie, Universit\"at Konstanz, 78457 Konstanz, Germany}
\author{Matthias Fuchs}
\affiliation{Fachbereich Physik, Universit\"at Konstanz, 78457 Konstanz, Germany}

\begin{abstract}
The smooth emergence of shear elasticity is an hallmark of the liquid to glass transition.
In a liquid, viscous stresses arise from local structural rearrangements. In the solid,  Eshelby has shown that stresses around an inclusion decay as  a power law $r^{-D}$, where $D$ is the dimension of the system. We study glass-forming hard sphere fluids by simulation and observe the emergence of the unscreened power-law Eshelby pattern in the stress correlations of the isotropic liquid state. By a detailed tensorial analysis, we show that the fluctuating force field, viz.~the divergence of the stress field, relaxes to zero with time in all states, while the shear stress correlations develop spatial power-law structures inside regions that grow with longitudinal and transverse sound speeds; we observe the predicted exponents  $r^{-D}$ and  $r^{-D-2}$.  In Brownian systems, shear stresses relax diffusively within these regions, with the diffusion coefficient determined by the shear modulus and the friction coefficient.
\end{abstract}

\maketitle

\section{\label{sec:Introduction} Introduction }

During the process of cooling, liquids start to show first traces of elasticity over increasing intermediate time windows. The phenomenon of viscoelasticity becomes especially strong in supercooled liquids when freezing can be prevented and the fluid approaches the transition to a glass \cite{Kob}. An observable reflecting this behavior is the time dependent shear modulus $G(t)$ \cite{Wagner2021,Larson}. It connects the stress response of a system to the rate of shear strain applied on the bulk (wave vector $q=0$) in the linear regime, and is a central quantity studied in rheology. Two extreme cases of the stress response are, on the one hand, $G(t)=\eta\delta(t)$ (with the Dirac delta $\delta(t)$) giving an ideal Newtonian fluid, and on the other hand $G(t)=G_{pl}= const.$ 
giving an elastic Hookean solid. In the former case, the shear stress $\sigma$ increases linearly with the shear rate $\dot\gamma$, viz.~$\sigma=\eta \dot{\gamma}$, and $\eta$ is identified as shear viscosity. In the latter case, the stress equals the constant shear modulus times the applied strain $\gamma$, $\sigma=G_{pl}\gamma$. According to Maxwell, viscosity and shear modulus are connected via the structural relaxation time $\tau$, viz.~$\eta= G_{pl} \tau$. Viscoelastic systems are in between those ideal cases and the time-dependence of the shear modulus $G(t)$ is important \cite{Nicolas2018}.
In the process of cooling a supercooled liquid, the $G(t)$ develops a shoulder over increasing time windows until ultimately $G(t)$ is non-zero at laboratory time-scales and the liquid freezes and becomes an amorphous solid. 

Via the linear response theory close to equilibrium, $G(t)$ is connected to the decay of thermal fluctuations of the (macroscopic) shear stress. This gives $G(t)$ a vivid interpretation.  On the other hand, the local  fluctuations of all elements of the stress tensor are the central object of this study.  In linear elasticity theory, describing the ideal Hookean solid as mentioned above, the space resolved stress correlations were calculated by Eshelby in order to determine the stress and strain fields around inclusions \cite{Eshelby}. The well known result is given by a characteristic octupolar angular dependence and an $r^{-D}$ distance dependence of the shear stress field induced by a force dipole \cite{Picard2004}. 

This leads to the question, if this stress pattern characteristic for elastic behavior can already be found in the shear stress correlation of an viscoelastic supercooled liquid. 
This asks if there are any correlation lengths in supercooled liquids beyond the trivial mean particle separation 
\cite{Todd2007,Puscasu2010,Tanaka_PRL_2009,Tong2020,JeppeDyre,Szamel2011,Flenner2015}. 
By theory, this was addressed using a generalized hydrodynamic approach, 
where the transition from local viscous to long-range elastic behavior was connected to momentum transport \cite{Maier2017,Manuel_JCP}.  A generalized Maxwell model has been derived, and already tested in simulations \cite{Rottler_JCP_2022}, finding that the Eshelby-pattern of elastic materials emerges in reciprocal space already in the supercooled liquid.
Our  contribution is concerned with the formation of the far field of the stress correlations in real space. We also address the connection between real and reciprocal space, and test the generalized hydrodynamic approach without Maxwell's approximations.
The emergence of elastic stress correlations in fluids originates from the coupling of stress fluctuations to momentum fluctuations. The development of a long range pattern can be argued as follows:
A shear stress fluctuation triggers a transverse momentum fluctuation, a shear wave. Due to the onset of elasticity in the supercooled liquid, this momentum fluctuation propagates and vice versa induces stress fluctuations on its trajectory. They are long lived as can be seen by the increasingly large relaxation times of the shear modulus. Since the momentum current is distributed over surfaces increasing with $r^{D-1}$ the displacements scale analogously. The stress is connected to the strain, i.e.~the derivative of these displacements, it therefore scales as $r^{-D}$ \cite{JeppeDyre}.

In this study we perform computer-simulations of hard spheres in three dimension (3D) and hard discs in two dimension (2D). The discussion in 2D is included because of the allowance for larger systems and a therefore more convincing measurement of the long ranged correlations. Furthermore the mathematical expressions are more handy, making technical aspects more transparent. \\
Since the stress correlation tensor is of fourth order it is non trivial to determine all its components in simulations. The computational and mathematical effort becomes manageable after a decomposition into tensseral harmonic tensors (real valued spherical harmonic tensors) as established for studies of the stress correlation tensor in zero temperature glasses \cite{Lemaitre_2015}.   \\
The shown simulations are space and time resolved, which gives the opportunity to directly observe the build-up of the various spatial patterns following the passing of longitudinal and transverse sound waves in Newtonian dynamics  \cite{Baschnagel_2018_SM}. We also perform Brownian dynamics simulations to study the  diffusive  shear stress transport in colloidal glasses \cite{Florian_EPL}. Hard sphere fluids are an established model for colloidal dispersions, where confocal microscopy can track particles and determine stresses from close contacts \cite{Lin2016}. The stress far fields have not been recorded yet in colloidal fluids, which we aim to overcome by developing fluorescently marked soft droplets as stress senors.

The paper is organized as follows. Section \ref{sec:simulation_prep} describes the preparation and the execution of the molecular dynamics (MD) simulations. How the stress fields are measured in these MD simulations is described in Sec.\,\ref{sec:stress_MD}, and Sec.~\ref{sec:tensorialAspects} recalls the symmetry based decomposition of the fourth rank tensor of stress correlation functions. There it will become clear, that in two dimension fewer independent functions arise than in three, which causes us to first present stress correlations in two dimensions, Sec.~\ref{sec:twoD}, before addressing three dimensions in Sec.~\ref{sec:threeD}. A discussion section including the presentation of stress sensor droplets  ends the manuscript.

\section{Simulation Preparation} \label{sec:simulation_prep}
We perform event driven simulations of hard particles interacting only when in contact via an elastic collision rule \cite{EDBD_algorithm}. To prevent crystallization we use a binary mixture of equal particle numbers $N_s/N_l=1$ and a diameter ratio of $d_l/d_s=1.4$, where $l$ and $s$ denote large and small spheres, respectively.
In two dimensions, we work at a density of $\phi=0.79$, slightly below the estimated glass transition of mode coupling theory (MCT) \cite{Goetzebook} at $\phi^c=0.7948$ \cite{WeysserHajnal2011,Fritschi2017}.  
In three dimension the density is $\phi=0.59$ at the MCT glass transition  \cite{vanMegen1991,Brambilla,Reinhardt2010}.
As seen in the various shear moduli below (Figs. \ref{fig:shearMod_N_2D}, \ref{fig:shearModulus_2D_BD} \ref{fig:shearModulus_Newton_3D}, \ref{fig:shearModulus_BD_3D.pdf}) the macroscopic stresses relax within the observation time.
Units are set by the diameter $d_l$, the thermal velocity $v_0$, and the thermal energy density $k_BTn$.

We also perform Brownian dynamics simulations based on the event driven algorithm. The damping is achieved by a Brownian thermostat re-drawing particle velocities from a Maxwell-Boltzmann distribution at fixed time-steps $\tau_b$.
The Brownian thermostat connects the friction to the Brownian time-step via $\zeta=2/\tau_b$ \cite{EDBD_algorithm}. We use $\tau_b=5\times 10^{-5}\,d_l^2/D_0$. With these technique, time-dependent shear moduli were already obtained in 2D \cite{Fritschi2017} and 3D \cite{Lange2009}, but not related to the other stress correlation functions.

\section{Stress in hard sphere simulations} \label{sec:stress_MD}
The stress $\tensor{\sigma}(\vec{r},t)$ describes a momentum current at time $t$ at location $\vec{r}$ and is therefore a second rank tensor \cite{HansenMcDonald}.
To extract the stress field in real space, the box of volume $V$ is subdivided into boxes of volume $V_B$, which have their center at  $\vec{r}_B$. The average over collisions happening in a time window $\Delta t = [t-\frac{\delta t}{2}, t+\frac{\delta t}{2}] $ around $t$ of length $\delta t$ at points in the box $V_B$ around $\vec{r}_B$ is taken to determine the stress tensor at time $t$ and position $\vec{r}_B$ 
 \begin{gather}
    \tensor{\sigma}(\vec{r}_B,t) = \frac{1}{V_B  \delta t}\sum_{\substack{\tau_c \in \Delta t; \, \vec{r}_c \in V_B}} \Delta\vec{r}^c\Delta\vec{p}^c. \label{eq:DefStress_sim_real}
\end{gather}
 Here $\Delta \vec{r}^c$ denotes the vector connecting the centers of the colliding spheres, $\Delta \vec{p}^c$ denotes the transferred momentum, $\tau_c$ the time of the collision and $\vec{r}_c$ its position.
 Note that the finite duration of our time-window, where many collisions are averaged, smoothens the short-time stress divergences expected for hard sphere interactions \cite{Dufty2002}. 
 We subdivide the simulation box in a grid width an edge length of 41 equally sized boxes in two and three dimensions, which determines the volume $V_B$. This leads in 2D to $L_B\approx 6.3\,d_l$ and in 3D to $L_b\approx 1.65 \, d_l$.
 
 We define the stress field in Fourier space as
 \begin{gather}
    \tensor{\tilde{\sigma}}(\vec{q},t) = \frac{1}{\delta t }\sum_{\tau_c\in  \Delta t} \, \Delta \vec{r}^c\,\Delta \vec{p}^c \, e^{i\,\vec{q}\cdot \vec{r}_c}, \label{eq:DefStress_sim_fourier}
\end{gather}
 where the sum runs over all collisions in the simulation box happening in the respective time window. \\

 The correlation functions in real and reciprocal space are then defined as 
 \begin{gather}
      \tensor{\sAC}(\vec{r}_B-\vec{r}_B', t-t') =  \mean{\tensor{\sigma}(\vec{r}_B, t)\;\tensor{\sigma}(\vec{r}_B', t') }, \\
       \tensor{\tilde{\sAC}}(\vec{q}, t-t') = \frac{1}{V} \mean{\tensor{\tilde{\sigma}}^*(\vec{q},t)\;  \tensor{\tilde{\sigma}}(\vec{q},t')}.
 \end{gather}
 
 Note that in real space (Eq.\,\ref{eq:DefStress_sim_real}), the collisions are first coarse grained over boxes, which then are correlated. In contrast in Fourier space  (Eq.~\ref{eq:DefStress_sim_fourier}), every collision individually contributes to the stress field, that is Fourier transformed. \\

 The shear modulus $G(t)$ follows then from the fluctuation dissipation theorem as
 \begin{gather}\label{eq:defShearmoulus_Sim}
     G(t) =\frac{1}{k_B T} \, \tilde{\sAC}_{xyxy}(\vec{q}= 0, t), 
 \end{gather}
i.e. the off-diagonal stress contributions of all collisions happening in the simulation box are summed over and auto-correlated in time. 
Throughout this work we assume stationary in time and are therefore free to choose $t'=0$.

An important aspect is collecting a large enough statistical data base since we aim for the, in general, small signals of the power law stress correlations at large distances. Thereto, it is important to use the isotropic symmetry of the viscoelastic fluid in order to  perform  angular averaging. The formalism that enables us to do so is described in the next section.

\section{Tensorial aspects of the stress-AC} \label{sec:tensorialAspects}
 The correlation function of the stress tensors is a tensor of fourth order with $D^4$ components in $D$ dimensions. This number can be reduced drastically by exploiting the symmetry of the liquid state and the properties of the stress tensor itself \cite{Lemaitre_2015,Lemaitre_2D}. In these seminal papers, the fourth order tensors of the stress correlations were expressed in spherical coordinates and linearly combined to give real valued spherical harmonic tensors (tesseral decomposition). This has three major advantages: i) it simplifies the identification of the non vanishing tensorial contributions, ii) the basis is orthogonal simplifying the algebra and iii) the individual tensor components are isotropic functions allowing angular averaging.  
These ideas have as well been employed in Ref.~\cite{Baschnagel_2023_PRE}, where spatially resolved stress correlations in glasses were investigated by simulations. \\

Next, the notation will be introduced that will be used thought the work.
We assume stationarity in time, and homogeneity and isotropy in space, so without loss of generality we can re-write the stress-autocorrelation tensor (SACT)  $\tensor{\sAC}$ with one time and one space argument
\begin{gather}
\tensor{\sAC}(\vec{r},t) = \mean{\tensor{\sigma}(\vec{r}-\vec{r}',t-t')\; \tensor{\sigma}  (\vec{r}',t') }. \label{eq:def1sAC}
\end{gather}
The system conserves angular momentum in every particle collision, so the $\tensor{\sigma}(\vec{r},t)$ is symmetric \cite{MartinParodiPershan}. \\
This implies the minor symmetries
\begin{gather}
    \sAC_{\abgd} = \sAC_{\alpha \beta \delta \gamma} =\sAC_{\beta \alpha \gamma \delta} \label{eq:minorSym}.
\end{gather}
Here and throughout, Greek indices label the spatial directions. 
Symmetry under time-inversion holds, which implies the major symmetry
\begin{gather}
    \sAC_{\alpha\beta \gamma\delta} = \sAC_{\gamma\delta \alpha \beta} \label{eq:majorSym}.
\end{gather}
Invariance under inversion and rotation means, that for every orthogonal matrix $\tensor{\mathrm{R}}$ it holds that 
\begin{gather}
   \tensor{\sAC}(\tensor{\mathrm{R}}\cdot \vec{r},t)= \tensor{\mathrm{R}}\,\tensor{\mathrm{R}}:\tensor{\sAC}(\vec{r},t):\tensor{\mathrm{R}}^T \tensor{\mathrm{R}}^T \label{eq:isotropyInvarianceTensor}
\end{gather}

The requirements on the tensor field given by Eqs.~\eqref{eq:minorSym}, \eqref{eq:majorSym}, and \eqref{eq:isotropyInvarianceTensor} constrain the SACT to an extent, where it can be written in terms of only 5 tensorial contributions $\tensor{\mathrm{Q}}^{(i)}(\uvec{r})$ with scalar valued pre-factors $q^{(i)}(r,t)$ that are isotropic functions of the scalar distance $r=|\vec{r}|$ only \cite{Lemaitre_2015}; here  $\uvec{r}=\vec{r}/r$ is a direction vector. 
This decomposition  reads 
\begin{gather}
    \tensor{\sAC}(\vec{r},t) = \sum_{i=1}^5  q^{(i)}(r,t) ~ \tensor{\mathrm{Q}}^{(i)}(\uvec{r}). 
\end{gather}
The explicit expressions of the tensors $\tensor{\mathrm{Q}}^{(i)}(\uvec{r})$ are given in three dimensions by Eqs.\,\eqref{eq:BasisLemaitre_3D} in App.~\ref{AppA3}.

The same decomposition is valid in Fourier space, where the basis tensors only change their dependencies from unit position vectors $\uvec{r}$ to unit wave vectors $\uvec{q}$ (where $\uvec{q}=\vec{q}/q$), i.e. $\tensor{\mathrm{Q}}^{(i)}(\uvec{q})$. Concomitantly,  the isotropic functions are dependent on the wavenumber $q$ and will be marked by a tilde, i.e. $\tilde{q}^{(i)}(q,t)$.
As Lemaître has emphasized and will be important later, the isotropic functions are not connected by a one dimensional Fourier transformation, i.e. $FT[q^{(i)}(r,t)] \neq \tilde{q}(q,t)$. Under Fourier transformation the tensorial structure and the spatial dimension of the problem leads to a mixing of the functions $q^{(i)}(r,t)$ and $\tilde{q}^{(i)}(q,t)$. 

The connection between the small-$\vec{q}$ structure of the correlations in reciprocal space and the far field spatial power laws depends on the dimension of space.
The same considerations in two dimension lead to only four tensorial contributions given by Eqs.\,\eqref{eq:BasisLemaitre_2D} in  App.~\ref{AppA2} \cite{Lemaitre_2D}. 
For ease of presentation, we therefore first discuss stresses in two dimension before moving to the three  dimensonal system of relevance to colloidal hard sphere suspensions.

The stress correlations in Cartesian basis are easily recovered from the isotropic functions $q^{(i)}(r,t)$.
For example in two dimensions, the  correlation function of an off-diagonal element reads 
\begin{gather}
    \sAC_{xyxy}(\vec{r},t) =   {q}^{(3)}(r,t)\, \left( 2 \,\hat{r}_x^2 \hat{r}_y^2 \right)+ {q}^{(4)}(r,t) \br{ \frac{1}{2} - 2 \hat{r}_x^2 \hat{r}_y^2 }  \label{eq:ssAC_2D}. 
\end{gather}
The corresponding expression in 3D will be given in Eq.~\eqref{eq:ssAC_3D_Fourier} below. 

The symmetry based decomposition of the stress correlations has a profound impact on the statistical averaging of the simulation data. As already mentioned, this is essential since we are concerned with the small signals of the power law correlations at large distances. The first aspect is the generally reduced number (five in $D=3$ and four in $D=2$) of independent functions $q^{(i)}$ and $\tilde{q}^{(i)}$ from the generally $D^4$ number of independent functions $\sAC_{\abgd}(\vec{r},t)$. But even more important is the isotropic symmetry of the functions $q^{(i)}(r,t)$ since then angular averaging becomes possible, which increases the statistics. Alongside it simplifies the presentation of the results, since the whole SACT can be represented by scalar valued functions with scalar dependencies.


 \section{Stress correlations in two dimensions}
\label{sec:twoD}
 
 \subsection{Shear modulus and zero wavevector limit}
 Similar to Eq.\,\eqref{eq:ssAC_2D} in real space, the corresponding shear stress auto-correlation in reciprocal space reads
\begin{gather}
   \tilde{\sAC}_{xyxy}(\vec{q},t) =  2\, \hat{q}_x^2 \hat{q}_y^2 \,\left(\tilde{q}^{(3)}(q,t)-\tilde{q}^{(4)}(q,t)\right)  + \frac{1}{2}\tilde{q}^{(4)}(q,t) \label{eq:ssAC_2D_fourier}. 
\end{gather}
This suggests the interpretation of $\tilde{q}^{(4)}(q,t)$ and $\tilde{q}^{(3)}(q,t)$ as the shear stress correlations along the axes and along the diagonals, respectively. If we take the $q\to 0$ limit for finite times, we recover the shear modulus because of isotropy. Therefore the limit $q\to 0$ needs to be approached independently of the direction, leading to the constraint $\lim_{q\to0} \left(\tilde{q}^{(3)}_{q,t}-\tilde{q}^{(4)}_{q,t}\right)=0$ at finite times. 
This consideration rules out any Eshelby power-law spatial dependence at finite times and distances, as will become clear later. It would lead to a non-analytic small $q$ limit.  
 To summarize, in the (isotropic) liquid the $q\to 0$ limit for finite times leads to  the shear modulus
 \begin{gather}\label{eq:shearMod_2D}
 \tilde{q}^{(3)}(0,t)=\tilde{q}^{(4)}(0,t)=2G(t)~. 
 \end{gather}

\subsection{Force correlations}
From the tensorial decomposition in reciprocal space, the correlation functions of the $\vec{q}$-dependent forces can be calculated. The force $\vec{F}$ is the divergence of the momentum current, i.e. $\tilde{\vec{F}}(\vec{q},t)=i \vec{q} \cdot  \tensor{\tilde{\sigma}} (\vec{q},t)$. 
Similar to velocity correlations, the force correlation tensor for isotropic systems can then be decomposed into longitudinal and transverse parts.
This yields the (rescaled by $q^{-2}$) force correlation function
\begin{gather}
    \tensor{Z}(\vec{q},t)= \uvec{q}\cdot \tilde{\tensor{\sAC}}(\vec{q},t)\cdot \uvec{q}  =  \uvec{q}\,\uvec{q} Z^\parallel(q,t) + ({\bf 1} - \uvec{q}\,\uvec{q}) \, Z^\perp(q,t) . \label{eq:force2D_decomp}
\end{gather}
Explicitly computing both divergences leads to expressions of the  force correlation functions in terms of the $\tilde{q}^{(i)}$ of the SACT.
This reads in two dimensions 
\begin{gather}\label{eq:forceCorr2D}
    Z^\parallel(q,t) = \left(\frac{1}{2}\tilde{q}^{(1)}(q,t) -\frac{1}{\sqrt{2}}\tilde{q}^{(2)}(q,t)  + \frac{1}{2}\tilde{q}^{(3)}(q,t) \right)  , \\
    Z^\perp(q,t)  =   \frac{1}{2} \tilde{q}^{(4)}(q,t)~. 
\end{gather}
The global longitudinal  modulus is the corresponding limit of the longitudinal force correlation function $K(t)=Z^\parallel(0,t)/k_B T$ and the shear modulus (Eq.~\ref{eq:shearMod_2D}) is connected to $G(t) = Z^\perp(0,t)/k_B T$. But again, these identifications are only valid for finite times. The general relations will be given in Eq.~\eqref{eq24}  below.

 \subsection{Fourier relations and far fields}

 The tesseral tensors in Fourier space $\tensor{\mathrm{Q}}^{(i)}(\hat{q})$ can be transformed to real space by inverse Fourier transformation; denoted $FT^{-1}$. For considering the spatial far field, integrals like the following arise (with Kronecker delta $\delta_{\alpha\beta}$ and $r>0$): 
 \begin{gather}\label{eq15}
\int \frac{\text{d}^2 \vec{q}}{4\pi^2} \hat{q}^2_\alpha \hat{q}^2_\beta e^{i\vec{q}\cdot\vec{r}} e^{-\epsilon\,q}\big|_{\epsilon=0} = \frac{1}{4\pi} \frac{1}{r^2} \left(5 \delta_{\alpha\beta} + 8 \hat{r}_\alpha^2\,\hat{r}_\beta^2 \right).
 \end{gather}
 Obviously, the right hand side can be decomposed into a power-law distance dependence multiplied by tesseral tensors in real space. Considering all tensors and collecting all terms, one finds the following relations \cite{Lemaitre_2D}: 
 \begin{gather}
 FT^{-1} \left[
 \begin{pmatrix}
        \tensor{\mathrm{Q}}_1(\uvec{q}) \\
        \tensor{\mathrm{Q}}_2(\uvec{q}) \\
        \tensor{\mathrm{Q}}_3(\uvec{q}) \\
        \tensor{\mathrm{Q}}_4(\uvec{q}) 
    \end{pmatrix} \right] 
    = \nonumber\\ 
    \frac{1}{2\pi r^2}\begin{pmatrix} 
   0  &0 & 0 & 0\\
   0 & -2 & 0 & 0 \\ 
   0 & 0 & 2 & -2\\ 
   0 &  0 & -2 & 2
    \end{pmatrix}   \begin{pmatrix}
        \tensor{\mathrm{Q}}_1(\uvec{r}) \\
        \tensor{\mathrm{Q}}_2(\uvec{r}) \\
        \tensor{\mathrm{Q}}_3(\uvec{r}) \\
        \tensor{\mathrm{Q}}_5(\uvec{r}) 
    \end{pmatrix},~\text{for}~r>0. \label{eq:InvFourierTensor_2D}
\end{gather}

The amplitudes of the far-fields in real space, viz.~$ {q}^{(i)}(r\to \infty,t) \propto r^{-D}$, can then be connected to the zero wavenumber limits, $q\to0$, of the isotropic functions in reciprocal space, i.e.~$\tilde{q}^{(i)}(0,t)$. These relations can be derived by
\begin{gather}
   \lim_{r\to \infty} \,  \tensor{\sAC}(\vec{r},t) =  \lim_{r\to \infty} \, FT^{-1}\left[ \tilde{\tensor{\sAC}}(\vec{q},t) \right] \nonumber \\
   = \sum_{i} \tilde{q}^{(i)}(0,t) \,FT^{-1} \left[ \tensor{\mathrm{Q}}_i(\uvec{q})  \right] \nonumber \\
   \overset{!}{=} \sum_i q^{(i)}(\infty,t) \tensor{\mathrm{Q}}_i(\uvec{r}) \label{eq:InvFourier_schematic}
\end{gather}
and since  the terms $FT^{-1} \left[ \tensor{\mathrm{Q}}_i(\uvec{q}) \right]$ are power laws following Eq.~\eqref{eq15}, 
the $\tilde{q}^{(i)}(0,t)$ determine their amplitudes. The exclamation mark marks that set of linear equations, where the identification of the pre-factors of the tensors $\tensor{\mathrm{Q}}_i(\uvec{r})$ can be performed.
The result then reads
\begin{gather}\label{eq:RealFourier_2D}
 \begin{pmatrix}
        {q}^{(1)}(\infty,t) \\
        {q}^{(2)}(\infty,t) \\
        {q}^{(3)}(\infty,t) \\
        {q}^{(4)}(\infty,t) \\
    \end{pmatrix}
   = \frac{1}{2\pi r^2} \begin{pmatrix} 
   0  &0 & 0 & 0\\
   0 & -2 & 0 & 0 \\ 
   0 & 0 & 2 & -2\\ 
   0 &  0 & -2 & 2
    \end{pmatrix} 
     \begin{pmatrix}
        \tilde{q}^{(1)}(0,t) \\
        \tilde{q}^{(2)}(0,t)  \\
        \tilde{q}^{(3)}(0,t)  \\
        \tilde{q}^{(4)}(0,t)  \\
    \end{pmatrix}. 
\end{gather}
This consideration of Fourier transformation rules alone yields the relation  $q^{(3)}(\infty,t)   = -q^{(4)}(\infty,t)$. The far field stress correlations have to obey this relation, which will be used as consistency check in the simulations.

\subsection{Force free solids and their fluid precursors}

\subsubsection{Inherent states as minima in  potential energy landscape}

In stable configurations at long times, the forces on each particle need to balance. Thus correlations of the force fields need to vanish, including in metastable glasses. This has been discussed in most details for inherent states \cite{Lemaitre_2015}.

Inherent states (IS) denote configurations of amorphous elastic solids at zero temperature, i.e. the potential energy minima of non-crystalline particle configurations. In these minima all particles are force-free, i.e. the divergence of the stress tensor vanishes. This means that the force correlations from Eq.\,\eqref{eq:forceCorr2D} vanish for long times in IS.
This is consistent with the results derived in Ref.~\cite{Lemaitre_2D} for IS
\begin{subequations}\label{eq:relations2DinQ}
\begin{gather}
  \tilde{q}(q\to 0,\infty)^{(1)}=\tilde{q}^{(2)}(q\to 0,\infty)/\sqrt{2}=\,\tilde{q}^{(3)}(q\to 0,\infty) \, \\
    \tilde{q}^{(4)}(q\to 0,\infty)=0. \label{eq:relations2DinQ_b}
\end{gather}
\end{subequations}
 If one uses the results of Eq.~\eqref{eq:relations2DinQ} in Eq.~\eqref{eq:RealFourier_2D}, one finds the IS relations in real space to read
 \begin{gather}
  q^{(3)}(\infty,t) = -q^{(2)}(\infty,t)/\sqrt{2} =  -q^{(4)}(\infty,t) \label{eq:Real_Power_2D_IS}.
 \end{gather}

\subsubsection{Force and stress correlations in the liquid}
\label{sec:IS_Liquid}

There appears to be an inconsistency in Eqs.~\eqref{eq:relations2DinQ} and \eqref{eq:shearMod_2D}. In the force free solid, the long time limit of the transverse force correlations vanishes for small wavevectors, $   \tilde{q}^{(4)}(q\to0,\infty)=0$ (Eq.~\ref{eq:relations2DinQ_b}), while the macroscopic limit at $q=0$ is identified as shear modulus $G(t)$, viz.~$G(t)=\tilde{q}^{(4)}(q=0,t)/2$ in Eq.~\eqref{eq:shearMod_2D}. The inconsistency becomes glaring when one recalls Maxwell's interpretation of the long time limit of the shear modulus as shear elastic constant of the solid, $G^{solid}(t\to \infty) = G_{pl}$; the plateau shear modulus also gets denoted as $\mu$, i.e.~one of the two Lamé constants of an isotropic solid \cite{Landau}.  \textit{Apparently, the limits do not interchange.}

The glassy solid at non-zero temperature was investigated in \cite{Baschnagel_2023_PRE}, where the shear-modulus decays onto the well defined plateau value, viz.~$G(t\to \infty) = G_{pl}$ holds. In this case it was possible to observe a well defined shear modulus, even though the force correlations were zero, as a solid state was studied. 
The liquid case considered in the present work is more subtle since the shear modulus is only approximately constant over a finite period of time before structural relaxation sets in. We will therefore carefully select a time-window where the shear modulus can be considered constant before the structural relaxation sets in; it will be marked by horizontal bars in Figs.~\ref{fig:shearMod_N_2D}, \ref{fig:shearModulus_2D_BD}, \ref{fig:shearModulus_Newton_3D}, and \ref{fig:shearModulus_BD_3D.pdf}.

The apparent inconsistency between  Eqs.~\eqref{eq:relations2DinQ} and \eqref{eq:shearMod_2D} is resolved via the proper identification of the intrinsic response functions of the viscoelastic fluid to applied inhomogeneous deformations. The SACT gives the response to an external velocity gradient, which induces stresses but  also momentum currents in the system. This coupling can be determined via the Zwanzig-Mori projection operator formalism \cite{Maier2017,Manuel_JCP,Florian_EPL}. The intrinsic response functions \cite{Martin}, which are not affected by the hydrodynamic momentum transport, then describe the stress responses to the internal velocity gradients 
\cite{Vogel2020}. These response function were identified as the Zwanzig-Mori memory kernels, well familiar from liquid theory \cite{HansenMcDonald}. The force auto-correlation functions depend  on the the history of the memory kernels, and the limits $q\to0$ and $t\to\infty$ do not commute \cite{Maier2017,Manuel_JCP}. For Newtonian dynamics, the   non-Markovian relations are (in an isothermal setting, neglecting heat transport) \cite{Evans1981,EvansMorriss}
\begin{align} \label{eq24}
    \partial_t\,Z^{i}(q,t) + \frac{q^2}{mn} \int_0^t\!\! dt'\; G^{i}(q,t-t')\, Z^{i}(q,t')  = \partial_t\, G^{i}(q,t)  .
\end{align}
The superscript $i$ stand for longitudinal ($\|$) and transverse ($\perp$) correlation functions, and the $G^i(q,t)$ are the corresponding Zwanzig-Mori memory kernels; see Refs.~\cite{Maier2017,Manuel_JCP,Florian_EPL} for their definitions. They
approach the moduli in the macroscopic limit, $G^\|(q,t)\to K(t)$ and $G^\perp(q,t)\to G(t)$ for $q\to0$. And for $t\to\infty$, the moduli approach the elastic constants in the solid limit (denoted by subscripts, i.e.~$G_{pl}$ and $K_{pl}=K^{solid}(t\to\infty)$). \\
While the transverse force correlation has already been given in Eq.~\eqref{eq:forceCorr2D},  the Zwanzig Mori decomposition gives \cite{Manuel_JCP} 
\begin{gather}\label{schermodul}
  G^\perp(q,t) =  \frac{1}{2} \tilde{q}^{(3)}(q,t) + 
  {\cal O}(q^4) \, , 
\end{gather}
in the incompressible limit,  which holds well for the hard sphere fluids at the considered densities.
These  functions will be discussed together with the numerical results in the following figures.

 \subsection{Molecular dynamics simulations}

 \begin{figure}[ht]
    \centering
    \includegraphics{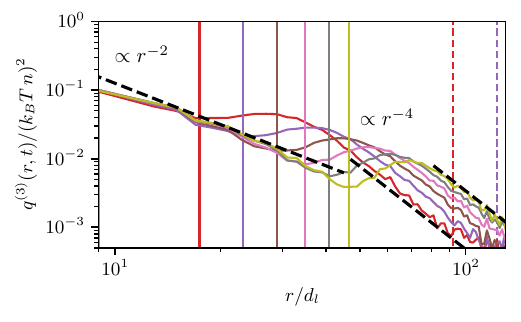}
    \caption{The build up of the stress correlation function $q^{(3)}(r,t)$, where different colors denote different times. The black dashed line for small distances describes a $r^{-2}$-power law, the two for lager distances describes a $r^{-4}$-power law. The power-laws emerge in the drag of the wave-fronts, since the shear wave triggers stress-fluctuations that decay on a much longer time-scale. The vertical solid lines correspond to $r=c_T\,t$ and the dashed vertical ones to $r=c_L\,t$ from Eq.\,\eqref{eq:sound_vels_2D_N}. The colors denote different times, matching the symbols in Figure \ref{fig:shearMod_N_2D}.  }
    \label{fig:wavefront_2D_N}
\end{figure}

\begin{figure}[ht]
    \centering
    \includegraphics{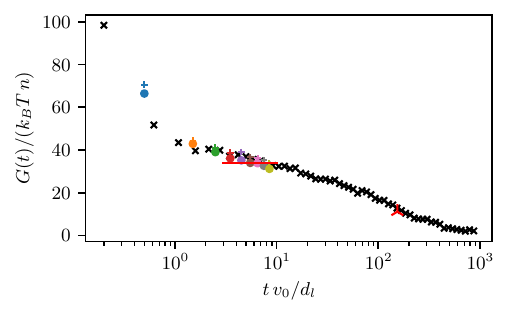}
    \caption{The bulk shear modulus $G(t)$ is shown by the black crosses. The colored crosses and dots denote the $q=0$ values of the $\tilde{q}^{(3)}$ and $\tilde{q}^{(4)}$ slopes from Fig.\,\ref{fig:q3_q4_F_N_time}. The colors as well coincide with the ones in Figure \ref{fig:wavefront_2D_N}. The triangle at long times denotes the time of Fig \ref{fig:IS_Comp_Newton_2D}. }
    \label{fig:shearMod_N_2D}
\end{figure}

 For Newtonian dynamics two mechanisms determine the structure of the stress correlation function. A fast longitudinal wave producing a $r^{-4}$ power-law followed by a shear wave, with a concomitant $r^{-2}$ algebraic decay \cite{Baschnagel_2018_SM}.
 For times at the plateau the shear wave propagates with $c_T=\sqrt{G_{pl}/(m\,n)}$ and $c_L = \sqrt{K_{pl}/(m\,n)}$. The transverse sound velocity is determined via the marked plateau in Fig.\,\ref{fig:shearMod_N_2D} (red bar), and the longitudinal one via the dispersion relation in Fig.\,\ref{fig:dispersionRelation_2D} in App.~\ref{AppB}. The results are
 \begin{gather}
     c_L = 31.51 \, v_0;  \qquad c_T = 5.47\, v_0, \label{eq:sound_vels_2D_N}
 \end{gather}
where $v_0$ is the thermal velocity. Figure \ref{fig:wavefront_2D_N} shows the built up of the power law as described in the introduction. The induced momentum flow is distributed over increasing surfaces (rings in two dimensions) scaling as $\propto 1/r$. The stress is given by its spatial derivative, leading to a $\propto 1/r^2$ law. 

In the shown time-frame of Figure \ref{fig:wavefront_2D_N}, the amplitude of the $r^{-2}$ algebraic law is almost constant. Only the distance over which it can be observed increases. Eq.~\eqref{eq:RealFourier_2D} and Eq.~\eqref{eq:shearMod_2D} together seem to suggest, that the power law amplitude vanishes. This would indeed be the case for $r\to \infty$ at finite times as assumed in Eq.~\eqref{eq:shearMod_2D}. But for the finite distances at finite times that are shown in Fig.~\ref{fig:wavefront_2D_N} the IS-relations approximately are valid on these length scales. Therefore we can use the IS-predictions Eq.~\eqref{eq:relations2DinQ}  in Eq.~\eqref{eq:RealFourier_2D}, suggesting that the real space power-law is only determined by $\tilde{q}^{(3)}(0,t)$. This is indeed the case, as the dashed black line for the $-2$ power law has the amplitude of $q^{(3)}(0,t=4\,d_l/v_0)/\pi$, i.e. the time of the purple slope. The time-points of Fig.\,\ref{fig:wavefront_2D_N} are roughly on the plateau in Fig.\,\ref{fig:shearMod_N_2D}, therefore having an approximately equal amplitude of the $-2$ power law. \\
 The amplitude of the $-4$ power-law shows a pronounced time-dependency that was calculated be $6k_B T  G_{pl}^2t^2/\pi n\,m$ in Ref.~\cite{Baschnagel_2018_SM}. For the smallest and largest times this slope is drawn, where the $G_{pl}$ estimate is denoted as the red bar in Fig.~\ref{fig:shearMod_N_2D}.

\begin{figure}[ht]
    \centering
    \includegraphics{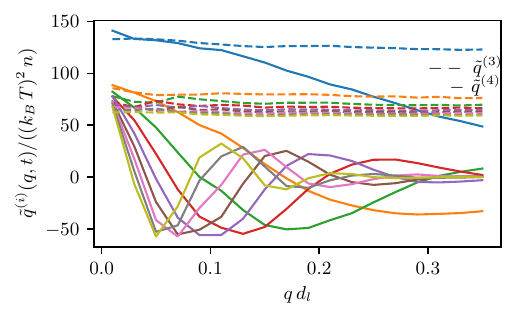}
    \caption{Dashed lines denote $\tilde{q}^{(3)}$ and full lines $\tilde{q}^{(4)}$. Different colors denote different times that can be read off from Fig.~\ref{fig:shearMod_N_2D}. It shows that $\tilde{q}^{(3)}_{0,t}=\tilde{q}^{(4)}_{0,t} $, while for $q\neq 0$ the function $\tilde{q}^{(4)}_{q,t}$ approaches zero for long times and $\tilde{q}^{(3)}_{q,t}$ stays rather constant, which is the IS-prediction from Eqs.\,\eqref{eq:relations2DinQ}. }
    \label{fig:q3_q4_F_N_time}
\end{figure}

Figure \ref{fig:q3_q4_F_N_time} shows the isotropic functions $\tilde{q}^{(3)}(q,t)$ and $\tilde{q}^{(4)}(q,t)$ with the same colors code the times as Figure \ref{fig:wavefront_2D_N}. One sees that both functions agree in the macroscopic limit,  $\tilde{q}^{(3)}_{0,t}=\tilde{q}^{(4)}_{0,t} $ for $q=0$ (Eq.~\eqref{eq:shearMod_2D}). On the other hand, the force-free prediction holds over successively larger ranges of $q$ for increasing time, $\tilde{q}^{(4)}_{q,\infty}=0 $  (Eq.~\eqref{eq:relations2DinQ_b}). While $\tilde{q}^{(3)}_{q,t}$ shows little variations with $q$.

The subtle time- and $q$-dependence can be explained by Eq.~\eqref{eq24}.
Entering a constant modulus  $G^{i}_{pl}(q)\approx\tilde{q}^{(3)}(q,t)/2$ into Eq.~\eqref{eq24}, one observes that the neglect of the integral requires smaller and smaller wavevectors for increasing time. Rather, the force correlation $Z^\perp(q,t)=\tilde{q}^{(4)}_{q,t}/2$ varies with the transverse sound velocity $\sqrt{G_{pl}/mn}.$ 
In the case of the solid, one finds that $\tilde{q}^{(3)}_{q,\infty}=const.$ while  $\tilde{q}^{(4)}_{q,\infty}=0$.  
This justifies the argumentation above, that local stress fluctuations approximately decay as slowly as the bulk shear-modulus.

Following Figure \ref{fig:q3_q4_F_N_time}  we observe that with increasing times the force-free relations Eq.\,\eqref{eq:relations2DinQ} hold for successively smaller wave vectors. This can be  observed in how $\tilde{q}^{(4)}$ vanishes for successively smaller $q$ with increasing time full-filling Eq.\,\eqref{eq:relations2DinQ_b}.  For $q=0$ it still approaches the value of $\tilde{q}^{(3)}$, which confirms Eq.\,\eqref{eq:shearMod_2D}. 

These results imply that the length scale, over which forces are relaxed, increases with increasing times, while elastic stresses remain in the solid. In isotropic solids this length scale is unbounded but in viscoelastic liquids the whole pattern decays with the decay of the shear modulus, that sets the amplitude of the power law and vanishes for long times in the liquid.
 
\begin{figure}[ht]
    \centering
    \includegraphics{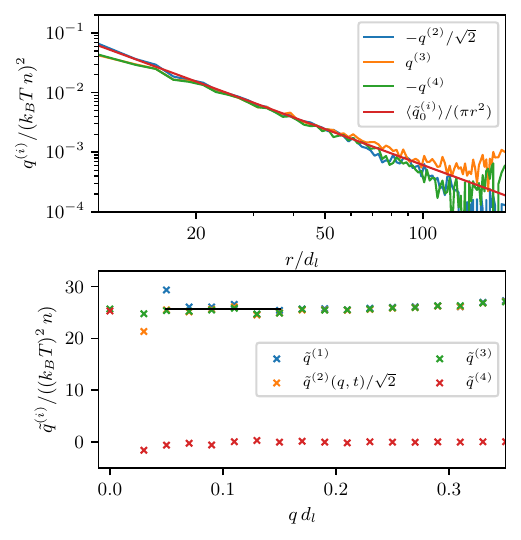}
    \caption{Top panel: For long times, the far fields show indeed the predicted power law $r^{-D}$ with amplitudes obeying the relations from Eq.\,\eqref{eq:Real_Power_2D_IS}. The red slope shows a power-law with an amplitude taken from the average value indicated by the black bar in the lower panel, i.e. the connection via Eq.\,\eqref{eq:RealFourier_2D}. \\ 
    Bottom panel:  In Fourier space, relation  Eq.\,\eqref{eq:relations2DinQ} is observed, as well as the fact that $\tilde{q}^{(3)}(0,t)=\tilde{q}^{(4)}(0,t)$ (smallest data point) from Eq.\eqref{eq:shearMod_2D}. }
    \label{fig:IS_Comp_Newton_2D}
\end{figure}

Figure \ref{fig:IS_Comp_Newton_2D} shows the isotropic functions in real and reciprocal space for much longer times (the red triangle in Fig.~\ref{fig:shearMod_N_2D}). As the bottom panel shows, for small but non-zero $q$'s, we can reproduce Eqs.~\eqref{eq:relations2DinQ}, i.e. the predictions of \cite{Lemaitre_2D}. The top panel shows the power laws in real space that coincide following Eq.~\eqref{eq:Real_Power_2D_IS}. The red bar is a power-law with the amplitude from the Fourier space values according to Eq.~\eqref{eq:RealFourier_2D}.

Ref.~\cite{Baschnagel_2023_PRE} is concerned with the solid case and finds that the only non-vanishing function is determined by Young's Modulus $E$. In our case, it is the shear modulus $G(t)$ that defines the $q=0$ limits. This difference originates in the different order of the limits: the limit $q\to 0$ at finite times (fluid), while results from Ref.~\cite{Baschnagel_2023_PRE} considers $t \to \infty$  at finite $q$ (solid).

 \begin{figure}[ht]
    \centering
    \includegraphics{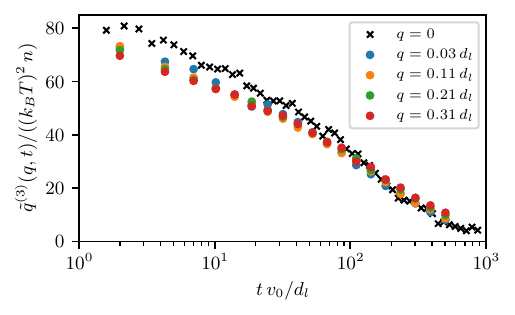}
    \caption{The final decay of $\tilde{q}^{(3)}(q,t)$ for the range of small wavevectors also covered in Fig.~\ref{fig:q3_q4_F_N_time}.  We do not observe a wave-vector dependency of the relaxation of $\tilde{q}^{3}(q,t)$. Follwoing Eq.~\eqref{schermodul} this implies a $q$-independent viscosity.} 
    \label{fig:q3_long_2D_newton}
\end{figure}

Extending the time window to the viscous regime, the wavevector dependent decay of the  correlation function of the fluctuating shear stresses can be tested. This gives insight into the wavevector dependence of the shear viscosity, which is given by the time-integral of the shear memory kernel \cite{HansenMcDonald}, viz.~
\begin{gather}\label{eq:viscosity}
\eta^\perp(q)=\int_0^\infty dt\, G^\perp(q,t) ~.  
\end{gather}
Figure \ref{fig:q3_long_2D_newton} reports the final decay of the $\tilde{q}^{(3)}(t)$ functions, which give the dominant contribution according to Eq.~\eqref{schermodul}. The lack of a $q$-dependence of the amplitude and of the viscous relaxation predicts that the shear viscosity does not depended on wave vector for the studied $q$-range. 

\subsection{Brownian dynamics simulations} \label{eq:SecBrownian2d}

In Ref.~\cite{Florian_EPL} a theory for shear stress correlations in Brownian systems was proposed. Even though frictional forces break the conservation of momentum that was used in the Newtonian theory \cite{Manuel_JCP}, it was possible to transfer the approach to Brownian systems. It was achieved by considering the transverse displacement field as a slow variable. The relation between force correlation functions and memory kernels from Eq.~\eqref{eq24} becomes overdamped:
\begin{align} \label{eq24B}
    \zeta\,Z^{i}(q,t) + \frac{q^2}{n} \int_0^t\!\! dt'\; G^{i}(q,t-t')\, Z^{i}(q,t')  = \zeta\, G^{i}(q,t)  ,
\end{align}
where $\zeta$ is the friction coefficient.
The resulting shear stress auto-correlation function in the over damped case reads 
\begin{gather}
    \sAC_{xyxy}(\vec{q},t) = G(t) \left[ 4\hat{q}_x^2\hat{q}_y^2 + (1-4\hat{q}_x^2\hat{q}_y^2)e^{-q^2\,D\,t} \right]  \label{eq:BD_FlorianEPL}.
\end{gather}
Here, the generalized hydrodynamic limit was taken. It corresponds to taking the $q\to0$ limit in the Zwanzig-Mori memory kernel, while its full time-dependence is kept, $G(t)=G_0^\perp(t)$. In Refs.~\cite{Maier2017,Manuel_JCP,Florian_EPL} an additional Maxwell approximation was performed, $G(t)\approx G^{gM}=G_{pl}\, e^{-t/\tau}+\Gamma \delta(t)$, which we can forego because $G(t)$ is known from the simulations.
The diffusion constant is connected to the plateau value of the shear modulus $G_{pl}$ and to the friction constant of the solvent $\zeta$, via $D=G_{pl}/(n\,\zeta)$ with $n$ being the number density. This result can easily be derived from Eq.~\eqref{eq24B}, and identifies the part of $ \sAC_{xyxy}(\vec{q},t)$ along the axis as force correlation $Z^\perp$.
This can easily be confirmed because along the axes Eq.~\eqref{eq:BD_FlorianEPL} simplifies to 
\begin{gather} \label{eq:diffusiveMode1}
     \sAC_{xyxy}(q\,\hat{x},t) = G(t)  e^{-q^2\,D\,t} ~,
\end{gather}
and from the tensorial decomposition Eq.\,\eqref{eq:ssAC_2D_fourier} we as well know 
\begin{gather}\label{eq:diffusiveMode2D}
    \sAC_{xyxy}(q \hat{x},t)= Z^\perp(q,t)=  \frac{1}{2} \tilde{q}^{(4)}(q,t)~.
\end{gather}
From Eq.\,\eqref{eq:diffusiveMode1} and Eq.\,\eqref{eq:diffusiveMode2D}, we can read off a Gaussian prediction for $\tilde{q}^{(4)}(q,t)$ for different times, where the two fit parameters of amplitude and variance give an estimate for the shear modulus $G(t)$ and for the diffusion coefficient $D$ respectively. 

To test these predictions, we perform Brownian dynamics simulations, and discuss the results in the following figures. 
Figure \ref{fig:c4Gaussian} shows the results, simulation data as solid and Gaussian fits as dashed lines. Similar to Figure \ref{fig:q3_q4_F_N_time} we can observe how the transverse force correlations decay to zero over larger length-scales for increasing times as predicted by Eqs.~\eqref{eq:relations2DinQ}.

\begin{figure}[ht]
    \centering
    \includegraphics{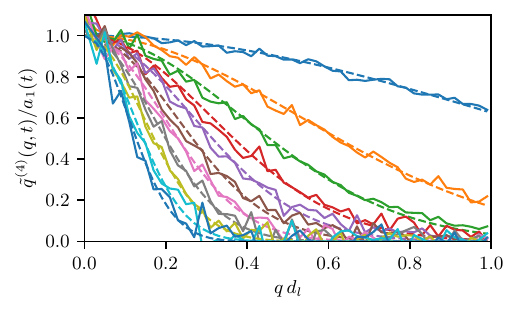}
    \caption{$\tilde{q}^{(4)}(q,t)$, where different colors show different times.  Following Eq.~\eqref{eq:diffusiveMode1}, the superimposed smooth lines show Gaussian fits of the form $a_1\, e^{-q^2\,b_1}$, $a_1$ giving an estimate for $G(t)$ and $b_1$ giving an estimate of $D\,t$, where $t$ is of course known. These relations are verified in Figs.~\ref{fig:shearModulus_2D_BD}, \ref{fig:fitDt}. }
    \label{fig:c4Gaussian}
\end{figure}
\begin{figure}[ht]
    \centering
    \includegraphics{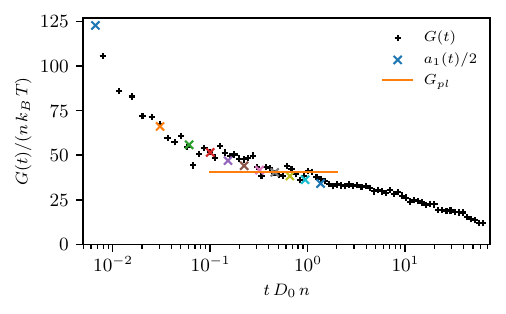}
    \caption{Shear modulus measured from collisions in the whole box (black) and from the amplitude parameters $a_1(t)$ fitted in Figure \ref{fig:c4Gaussian} (colors), which are connected via Eqs.\,(\ref{eq:diffusiveMode1}, \ref{eq:diffusiveMode2D}). We consider $G(t)$ to have an approximate plateau at the last seven data-points, and averaging over these data points results in the value of the orange bar, estimating $G_{pl}$. }
    \label{fig:shearModulus_2D_BD}
\end{figure}
\begin{figure}[ht]
    \centering
    \includegraphics{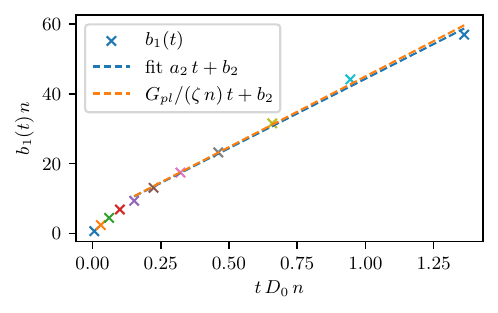}
    \caption{The crosses show the diffusive exponentials from the similarly colored slopes in Fig.\,\ref{fig:c4Gaussian}. The blue line shows a linear fit through these data points. The orange line gives the theory predictions, where $\zeta$ comes from the Brownian thermostat and $G_{pl}$ is taken from Fig.\,\ref{fig:shearModulus_2D_BD}. }
    \label{fig:fitDt}
\end{figure}

As an consistency check, Figure \ref{fig:shearModulus_2D_BD} shows $G(t)$ measured in a conventional way, Eq.\,\eqref{eq:defShearmoulus_Sim}, in black, while the colored scatters are the amplitude fit parameters $a_1(t)$ from Figure \ref{fig:c4Gaussian}. The orange bar denotes the mean over these points where the shear modulus $G(t)$ was assumed to have plateaued.
Figure \ref{fig:fitDt} shows the fit parameters $b_1(t)$ from Figure \ref{fig:c4Gaussian}. Since we know $G_{pl}$ from figure \ref{fig:shearModulus_2D_BD} and $\zeta$ from $\tau_b$, we can predict the diffusion. This prediction is plotted by the orange slope in Fig.~\ref{fig:fitDt}. For a comparison the blue one is just a linear fit, showing the good agreement with the orange slope. The linear fit has an off-set parameter $b_2$ that is as well used for the orange dashed slope. This terms accounts for non-diffusive effects of the shear mode for short times, where the modulus has not plateaued. The agreement with a linear fit is convincing validating the prediction that the correlations propagate diffusively in the regime of the plateau of $G(t)$. \\

\begin{figure}
    \centering
    \includegraphics{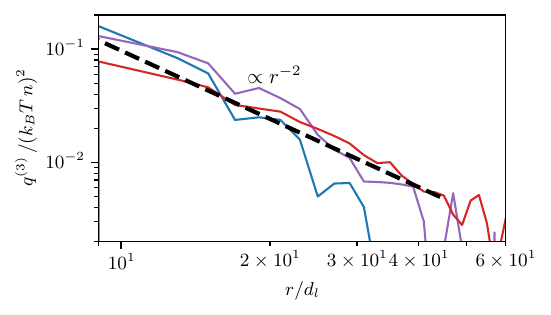}
    \caption{$q^{(3)}(r,t)$ in Brownian dynamics at $t\,D_0\, n = 0.33;\, 1.5;\,12.2$. It was possible to resolve the shear stress correlations up to a distance of roughly $r=50\,d_l$. }
    \label{fig:q3_wavefront_BD_2D}
\end{figure}

To this point the investigation of the overdamped system was concerned with the diffusive propagation of the mode. Figure \ref{fig:q3_wavefront_BD_2D} is concerned with the long ranged correlations that are as well present in the Brownian system. It shows the shear stress correlation along the diagonals, i.e. $q^{(3)}$ for three different times: $t\,D_0\, n = 0.33;\, 1.5;\,12.2$.  In contrast to the Newtonian dynamics the statistics is less accurate. The as well expected $r^{-4}$ behavior could not be resolved. The dashed line denotes the $r^{-2}$ power law with an amplitude given by the orange $G_{pl}$ estimate from Figure \ref{fig:shearModulus_2D_BD}. 
Since the tensorial decomposition is independent of the dynamics those are as well connected via Eqs.~(\ref{eq:shearMod_2D}, \ref{eq:RealFourier_2D}).

\section{Three Dimensions} \label{sec:threeD}
We repeat the previous analysis concerning two dimensional systems, but now in three dimensions. In three dimensions the decomposition consists of five functions $\tilde{q}^{(i)}$ connected to the tensors Eqs.~\eqref{eq:BasisLemaitre_3D} \cite{Lemaitre_JCP_2018}.

For instance the shear stress component reads in the three dimensional tesseral decomposition:
    \begin{gather}
    \sAC_{xyxy}(\vec{q},t)=  \frac{1}{2\sqrt{2}} \, \tilde{q}^{(4)}(q,t) + \nonumber \\  \hat{q}_x^2\, \hat{q}_y^2 \,\left(\frac{3}{2}\tilde{q}^{(3)}(q,t)   +\frac{1}{2\sqrt{2}}\,  \tilde{q}^{(5)}(q,t) - \sqrt{2}  \tilde{q}^{(4)}(q,t)  \right)   
    \label{eq:ssAC_3D_Fourier}.
\end{gather}
The anisotropic second term is connected to an $r^{-3}$ power law that vanishes for large distances at finite times. Nevertheless Eq.~\eqref{eq:defShearmoulus_Sim} needs to hold, which therefore yields a relation of the shear modulus to the $q=0$ value of $\tilde{q}^{(4)}(q,t)$
\begin{gather}
   G(t) =  \frac{1}{2\sqrt{2}} \tilde{q}^{(4)}(0,t).
\end{gather}

Along the diagonal in the $xy$-plane the corresponding real space decomposition in Eq.~\eqref{eq:ssAC_3D_Fourier} reads
 \begin{gather} \label{eq:Cxyxy_diag}
    \sAC_{xyxy}(r^{diag},t)=     \frac{3}{8}{q}^{(3)}(r,t) +\frac{1}{8\sqrt{2}}\,  {q}^{(5)}(r,t)  
\end{gather}

\subsection{Precursors of IS in the liquid}\label{sec:IS_liquid_3D}

The force correlation in three dimensions read (cf. Eq.\, \eqref{eq:force2D_decomp}):
\begin{subequations}
\begin{gather}
    Z^\parallel(q,t) = \left(\frac{1}{3}\tilde{q}^{(1)}(q,t) -\frac{2}{3}\tilde{q}^{(2)}(q,t)  + \frac{2}{3}\tilde{q}^{(3)}(q,t) \right) , \\
    Z^\perp(q,t) =   \tilde{q}^{(4)}(q,t) ~. \label{eq:transF_3D}
\end{gather}
\label{eq:forceCorr3D}
\end{subequations}
Confirming the IS calculations \cite{Lemaitre_JCP_2018} that yield 
\begin{subequations}
  \begin{gather}
   \tilde{q}^{(1)}(0,\infty) = \tilde{q}^{(2)}(0,\infty) = 2\,\tilde{q}^{(3)}(0,\infty)~,\\
    \tilde{q}^{(4)}(0,\infty) = 0~.
\end{gather}  \label{eq:IS_relations_3D}
\end{subequations}
Therefore only two independent functions determine the IS.

From the ZM-decomposition in the incompressible limit the shear modulus can be connected to the reciprocal version of Eq.~\eqref{eq:Cxyxy_diag}, which follows the analogous considerations as Eq.~\eqref{schermodul} in 2D but in 3D
\begin{gather}\label{eq:schermodul3D}
    G^{\perp}(q,t) =\frac{3}{8}\tilde{q}^{(3)}(q,t) +\frac{1}{8\sqrt{2}}\,  \tilde{q}^{(5)}(q,t)+\mathcal{O}(q^4) .
\end{gather}

The same considerations leading to Eq.~\eqref{eq:InvFourier_schematic} in three dimensions, yield  the connection of the power-law amplitudes $q^{(i)}(\infty,t)$ to the $q\to 0$ limits  in Fourier space  $\tilde{q}^{(i)}(q\to0)=\tilde{q}^{(i)}_0$ 
\begin{gather} \label{eq:Fourier_3D}
   \begin{pmatrix}
        {q}^{(1)}_{\infty,t} \\
        {q}^{(2)}_{\infty,t} \\
        {q}^{(3)}_{\infty,t} \\
        {q}^{(4)}_{\infty,t} \\
        {q}^{(5)}_{\infty,t} \\
    \end{pmatrix} = \frac{3}{4\,\pi r^3} \begin{pmatrix}
    0 & 0 & 0 & 0 & 0 \\
        0&-1 & 0 & 0 & 0 \\
         0&0 & 1 & -\sqrt{2} & \frac{1}{\sqrt{2}} \\
          0&0 & -\sqrt{2} & 1 & 0 \\
           0&0 & \frac{1}{\sqrt{2}} & 0 & -\frac{1}{2} \\
    \end{pmatrix} \begin{pmatrix}
        {\tilde{q}}^{(1)}_{0,t} \\
        {\tilde{q}}^{(2)}_{0,t} \\
        {\tilde{q}}^{(3)}_{0,t} \\
        {\tilde{q}}^{(4)}_{0,t} \\
        {\tilde{q}}^{(5)}_{0,t} \\
    \end{pmatrix} ~.
\end{gather}
This is the three dimensional analog of Eq.~\eqref{eq:RealFourier_2D}.

As expected, the pressure correlation is not long ranged which can be read off directly from $q_{\infty,t}^{(1)}=0$.
The lines are not linearly independent, from which the relation follows
\begin{gather}\label{eq:relations3DinR}
     q^{(3)}(\infty,t) = -\sqrt{2} \left( q^{(4)}(\infty,t) + q^{(5)}(\infty,t) \right) . 
\end{gather} 
 If one uses the IS results Eq.~\eqref{eq:IS_relations_3D}  in Eq.~\eqref{eq:Fourier_3D} another relation for the real space power laws emerges
\begin{gather}\label{eq:Relation2_3D}
    q^{(2)}(\infty,t) = \sqrt{2}\,q^{(4)}(\infty,t)~. 
\end{gather}

\subsection{Molecular dynamics simulations}
\begin{figure}[ht]
    \centering
    \includegraphics{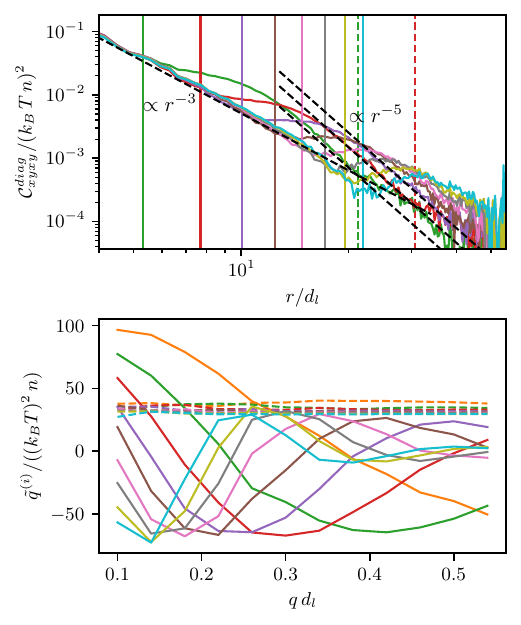}
    \caption{Top: the evolution of the $\sAC_{xyxy}$ shear stress correlation along the diagonals. The black dashed line at shorter distances denotes the $r^{-3}$ power law with a constant amplitude. The black dashed lines at larger distances denote the $r^{-5}$-power law for the three earliest times, which has a time dependent amplitude cf.~Eq.~\eqref{eq:r5_power}. Continuous vertical lines denote the expected position of the transverse wavefront, dashed vertical lines the position of the longitudinal wave. Bottom: solid lines show $\tilde{q}^{4}$,  denoting transverse force correlations that vanish. Dashed lines show the combination of $\tilde{q}^{(i)}$ given in Eq.~\eqref{eq:schermodul3D}, verifying the $q$-independence of the shear modulus. The colors top and bottom match the times shown in Fig.~\ref{fig:shearModulus_Newton_3D}. }
    \label{fig:wavefront_newton_3d}
\end{figure}

Figure \ref{fig:wavefront_newton_3d} shows the evolution of the shear stress correlations as the two wave fronts travel through the system. The simulation was run with Newtonian dynamics, allowing the appearance of oscillating sound waves as seen in the bottom panel. Similar to two dimensions the momentum current is distributed over spherical surfaces scaling as $\propto r^{-2}$, which leads to a stress tensor scaling proportional to its derivative $\propto r^{-3}$. The top panel shows this build up in real space, where the black bar is a guide to the eye. Furthermore we observe a $r^{-5}$ power law in the drag of a longitudinal soundwave. 
The vertical solid lines denote the position of the shear wave front, calculated from the transverse sound velocity given by the plateau value of the shear modulus Fig.~\ref{fig:shearModulus_Newton_3D} denoted by the horizontal bar. The dashed vertical lines denote the expected position of the soundwave front from the longitudinal sound velocity.
The different times correspond to similarly colored dots in Fig.~\ref{fig:shearModulus_Newton_3D}. 
The $r^{-5}$ amplitude can be calculated from ZM-considerations on the plateau to read 
\begin{gather}\label{eq:r5_power}
    \mathcal{C}_{xyxy}^{diag}(t) =  \frac{57}{8\,\pi}\,\frac{k_B T}{m\,n} \,G_{pl}^2\,t^2 \,r^{-5} ~.
\end{gather}
Most interestingly this power law amplitude increases in the time-window on the plateau. Fig.~\ref{fig:wavefront_newton_3d} shows the explicit results of Eq.\,\eqref{eq:r5_power} for the earliest three slopes, i.e. green, red and purple as black dashed lines. Again $G_{pl}$ was the estimate from the red bar in Fig.\,\ref{fig:shearModulus_Newton_3D}. For the green slope it is even possible to observe how the curve deviates from the $r^{-5}$ law at the position of the sound wave front (green dashed vertical line).

The bottom panel shows $\tilde{q}^{(4)}$ that is proportional to the transverse force correlations, cf. Eq.~\eqref{eq:transF_3D}. Therefore Eq.~\eqref{eq24} describes the behavior and simulations indeed show the expected waves. Also, the $\tilde{q}^{(4)}$ vanishes for smaller wave vectors as time increases. The dashed lines show the two $\tilde{q}^{(i)}$ of the rhs. of  Eq.~\eqref{eq:schermodul3D}. Indeed the $q$-dependence can be neglected as proposed by the ZM-calculations. 

\begin{figure}[ht]
    \centering
    \includegraphics{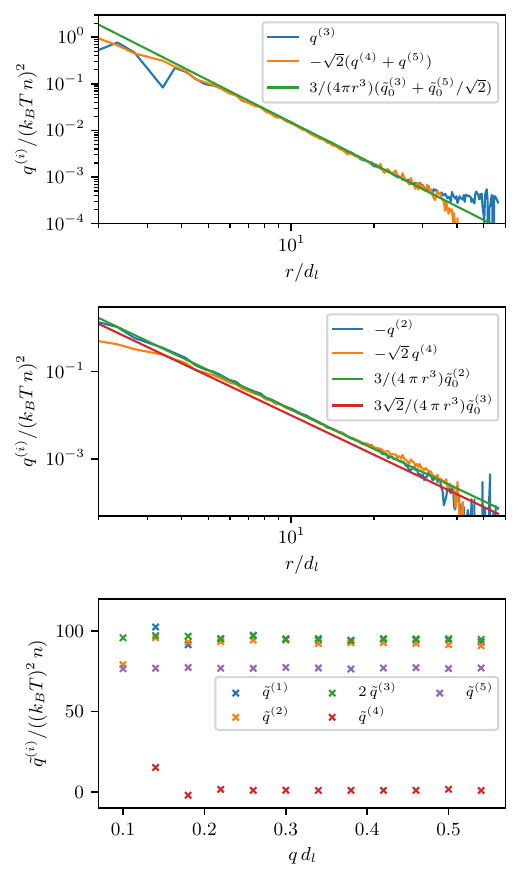}
    \caption{ The SACF functions at $t=15\, v_0/d_l$ and Newtonian dynamics. Top: relation Eq.~\eqref{eq:relations3DinR} is confirmed as well as the power-law amplitudes predicted by Fourier transformation via Eq.\,\eqref{eq:Fourier_3D}. Mid: Relation Eq.~\eqref{eq:Relation2_3D} is confirmed, as well as the corresponding part of Eq.~\eqref{eq:Fourier_3D}. Bottom: the relations from the IS predictions in Eqs.~\eqref{eq:IS_relations_3D} are confirmed well. }
    \label{fig:IS_comp_Newton_3D}
\end{figure}

\begin{figure}[ht]
    \centering
    \includegraphics{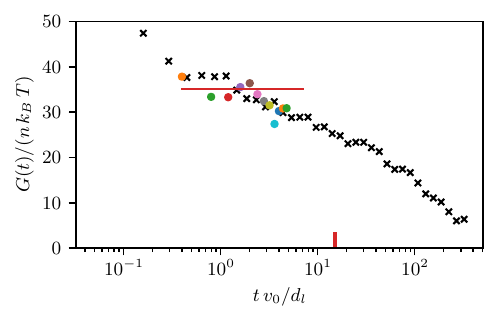}
    \caption{Shear modulus in Newtonian dynamics in 3D as black crosses. Circles denote the small wave-vector $q_{min}$ values of the similarly colored dashed lines in the bottom panel of Fig.~\ref{fig:wavefront_newton_3d}. The dots denote the times of the similarly colored slopes of top and bottom panel in Fig.~\ref{fig:wavefront_newton_3d}. The red vertical bar at $t=15\,v_0/d_l$ denotes the time of Fig.~\ref{fig:IS_comp_Newton_3D}. The red horizontal bar denotes the estimate of $G_{pl}$, which determines the transverse sound velocity.  }
    \label{fig:shearModulus_Newton_3D}
\end{figure}

The precursors of the IS relations in the liquid given in  Sec.~\ref{sec:IS_liquid_3D} can be observed in Figure \ref{fig:IS_comp_Newton_3D}. It shows the SACT at $t=15\,d_l/v_0$,  marked by the red vertical bar in Fig.~\ref{fig:shearModulus_Newton_3D}. The top panel shows real space measurements of the $q^{(i)}$ in a combination so that the far field power laws should coincide because of Eq.~\eqref{eq:relations3DinR}. Following Eq.~\eqref{eq:Fourier_3D} the far field of $q^{(3)}$ is determined by an $r^{-3}$ power-law with an amplitude given by a combination of several  $\tilde{q}^{(i)}(0,t)$. This is shown by the green line. 

The second relation Eq.~\eqref{eq:Relation2_3D} is shown in the center panel, together with the corresponding power laws determined by the Fourier space values again via Eq.~\eqref{eq:Fourier_3D}. 

The bottom panel shows that the IS relations in Fourier space Eqs.~\eqref{eq:IS_relations_3D} are present in the liquid state already. Nevertheless for the smallest $q$-values the forces seem to be not fully relaxed, giving rise to small deviations from the IS predictions at small $q$.

\begin{figure}
    \centering
    \includegraphics{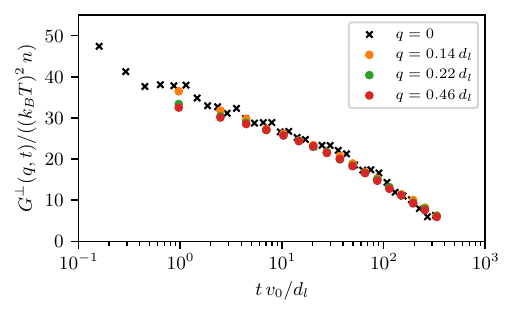}
    \caption{The three dimensional analog to Fig.~\ref{fig:q3_long_2D_newton}. Via Eq.~\eqref{eq:schermodul3D} the $\tilde{q}^{(i)}$ are connected to the generalized shear modulus $G^\perp(q,t)$. No length scale dependent relaxation is observed, leading to a $q$-independent viscosity according to Eq.~\eqref{eq:viscosity}.  }
    \label{fig:N_3D_q3_longRelax}
\end{figure}

Figure \ref{fig:N_3D_q3_longRelax} shows the generalized shear modulus $G^{\perp}(q,t)$ that can be calculated from the $\tilde{q}^{(i)}$ via Eq.~\eqref{eq:schermodul3D}. No legth scale dependent relaxation is observed, as already anticipated in the bottom panel of Fig.~\ref{fig:wavefront_newton_3d}. 

\subsection{Brownian dynamics simulations}

\begin{figure}[h]
\centering
\includegraphics[scale=1]{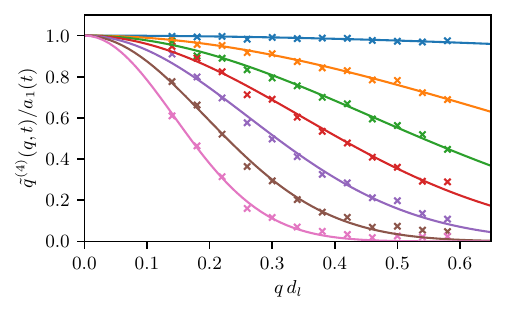}
\caption{Fit of the form $a_1\,e^{-q^2\,b_1}$ to $\tilde{q}^{(4)}(q,t)$ where different colors denote different times. Via Eq.~\eqref{eq:Diffusion_theroy_3d} we can connect $a_1(t)$ to the shear modulus (cf.~Fig.~\ref{fig:shearModulus_BD_3D.pdf}) and $b_1(t)=D\,t$ to diffusive dynamics (cf.~Fig.~\ref{fig:LinearFit_Diffusions_3d}). }
\label{fig:GaussianFits}
\end{figure}

\begin{figure}[h]
\centering
\includegraphics[scale=1]{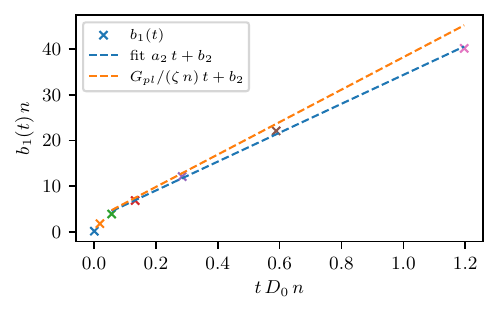}
\caption{ Crosses denote fit parameters $b_1(t)$ from Figure \ref{fig:GaussianFits} to every slope. The blue dashed line denotes a linear Fit with an offset $b_2$ accounting for non-diffusive short time effects. The orange dashed line denotes the theory prediction Eq.\eqref{eq:Diffusion_theroy_3d}, where the non-diffusive parameter $b_2$ was added. }
\label{fig:LinearFit_Diffusions_3d}
\end{figure}

\begin{figure}
\centering
\includegraphics[scale=1]{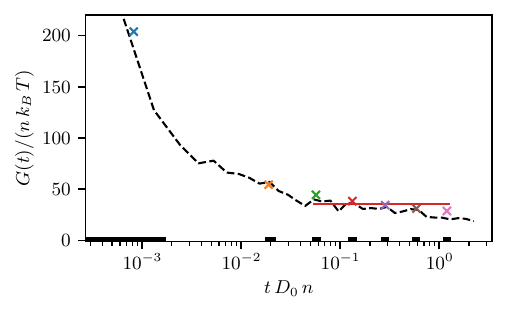}
\caption{From Eq.~\eqref{eq:Diffusion_theroy_3d} we can connect the amplitudes of the Gaussian fits in Fig.\,\ref{fig:GaussianFits} to the shear modulus. The black dashed line is calculated in the conventional way of correlating the stresses in the whole simulation box following Eq.~\eqref{eq:defShearmoulus_Sim}.  }
\label{fig:shearModulus_BD_3D.pdf}
\end{figure}

We perform the equivalent analysis of the Brownian dynamics in three dimensions as was already performed in two dimensions in Sec.\ref{eq:SecBrownian2d}. \\
By again considering only the directions along the axis in Eq.\,\eqref{eq:ssAC_3D_Fourier} (i.e. the isotropic part) from the ZM-projection operator formalism we find the connection to theory predictions of Ref.~\cite{Florian_EPL} to read
\begin{gather}
  \frac{1}{2\sqrt{2}}  \tilde{q}^{(4)}(q,t) = G(t)\, e^{-q^2\,D t } \label{eq:Diffusion_theroy_3d}.
\end{gather}
Analogous to Fig.~\ref{fig:c4Gaussian} in two dimensions, the theory in three dimensions  predicts a diffusive slope for $\tilde{q}^{(4)}(q,t)$ 
in three dimensions. This is shown in Figure \ref{fig:GaussianFits}.
The different colors in Fig.~\ref{fig:GaussianFits} are connected to times of the similarly colored data points in Fig.~\ref{fig:shearModulus_BD_3D.pdf}, which shows the shear modulus. The red bar denotes the estimate of $G_{pl}$. The time dependent variances of the Gaussian fits, denoted as $b_1(t)$ are shown by the scatters in Figure \ref{fig:LinearFit_Diffusions_3d} together with a fit (blue) through the data points where we consider the shear modulus to have plateaued. The agreement with a linear increase proves the diffusive propagation. The orange bar represents the theory prediction of the diffusive mode with a diffusion coefficient given by $D=G_{pl}/(\zeta n)$.

\section{Discussion}
In agreement with predictions from
non-Markovian Langevin equations approaches derived using  Zwanzig-Mori projection operators \cite{Maier2017,Manuel_JCP,Florian_EPL}, we could observe the emergence of long-ranged stress correlations in simulations of dense hard sphere liquids. These precursors of the Hookean elastic solid, which build up continuously in the supercooled liquid, result from the conservation law of momentum transport. While in a fluid, shear momentum diffuses, in the elastic solid it propagates via elastic waves. This crossover affects all elements of the tensor of stress correlations, which additionally contains the complexity that the long-time states have to be force free. In Brownian fluids, the analogous crossover is observed, with the difference that the stresses that originate from the interparticle interactions diffuse in the solid state.  We used the formalism of spherical tensors introduced in \cite{Lemaitre_2015} for mechanically rigid and time-stationary (inherent) states to analyze the stress correlations. This formalism proved itself also helpful in the time depended analysis of the fluid as performed in this contribution. 

Within the size of the simulation boxes in two and three dimensions convincing power-law correlations are observed as shown in Figs.  \ref{fig:wavefront_2D_N} and \ref{fig:wavefront_newton_3d}. Exponential screening as proposed in \cite{JeppeDyre} was not observed within our scope. Ref.~\cite{Tong2020} reports large non-zero average forces, which we do not find. Rather  our finding of spatio-temporal relaxation to a homogeneous  force-free state in the liquid  seems to align very well with the predictions of the stable, force-free system as considered in the IS approach. The necessary generalization to the viscoelastic liquid state is provided by the non-Markovian Langevin approach following Zwanzig and Mori.

The question about possible non-local contributions in the shear viscosity has been raised. 
While appreciable wavevector dependences have been observed in simulations of supercooled fluids, including of polymeric ones 
\cite{Todd2007,Puscasu2010,Tanaka_PRL_2009,Tong2020}, we find the (generalized) shear modulus and its viscous relaxation to be $q$-independent to a very good approximation. This difference warrants additional studies. 

\begin{figure}
\centering
\includegraphics[width=0.45\textwidth]{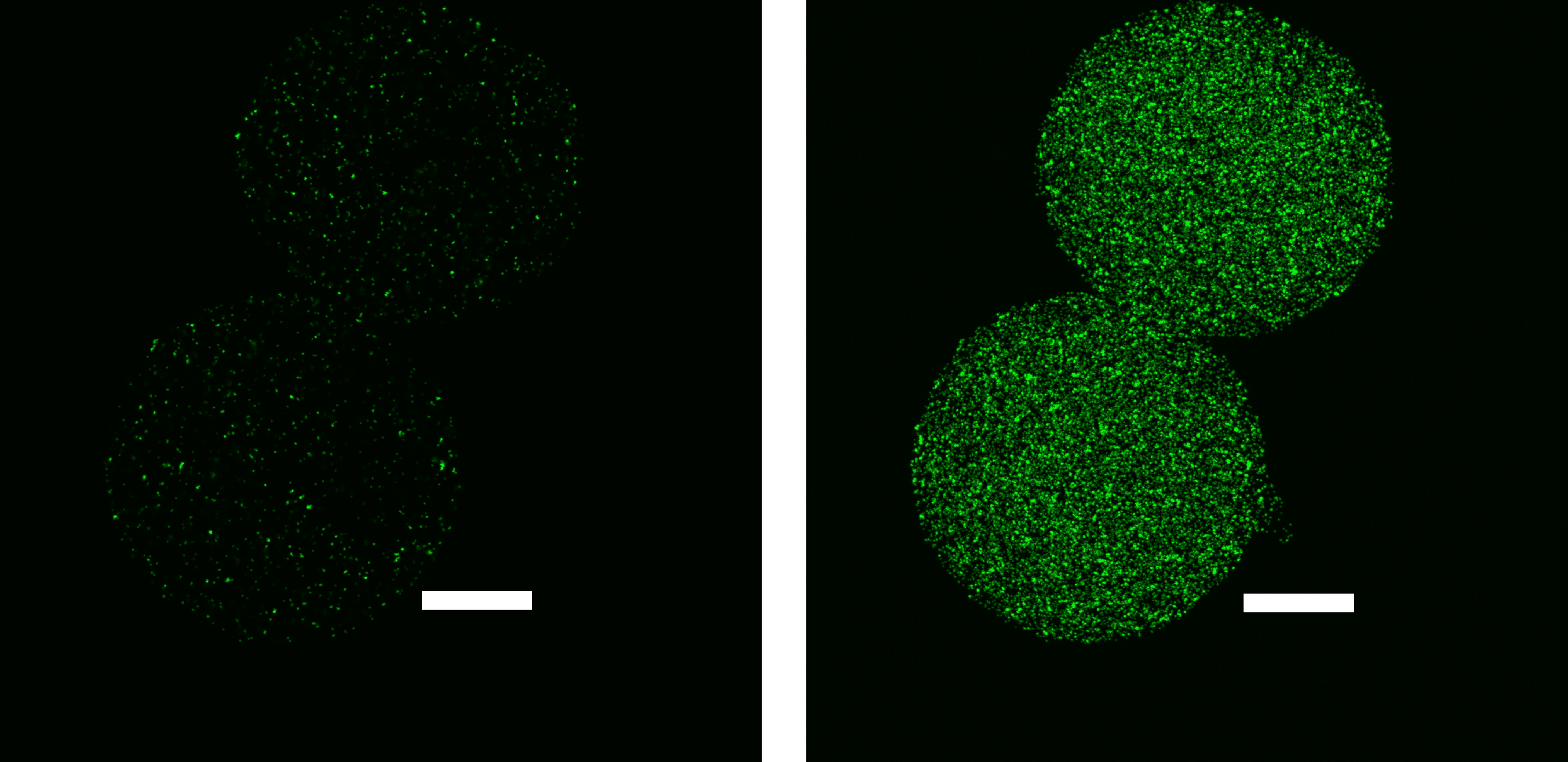}
\caption{Confocal fluorescence images of soft alginate force probe particles. Stress fields in the surrounding viscoelastic fluid lead to a deformation of the soft particles that could be detected as position changes of the fluorescent markers inside probe particle. Left: Scan in a single plane, right:  projection of all image planes. Scale bars: 30 $\mu m$}
\label{fig:particles}
\end{figure}

To test our predictions experimentally, we prepared probe particles serving as sensors for shear stress. The particles consist of  large, deformable alginate microgel droplets that contain hundreds of small fluorescent polymer beads (Fig.~\ref{fig:particles}). The diameter of the microgel shell is $\approx 100$\,$\mu$m, whereas the fluorescent beads have diameters of $200$\,nm. Particles of this type have previously been used as force sensors in biologial tissues \cite{Campas2014, Mohagheghian2018}. For the experiments, probe particles immersed in a dispersion of hard spheres allow recording the 3D positions of the fluorescent beads using confocal fluorescence microscopy. Spatial positioning accuracies of 20\,nm in xy and 50\,nm in the z-direction are commonly achieved and can be recorded over mesoscopic distances. Shear stress in the fluid leads to a deformation of the probe particles which can be analyzed as a change in the position of the fluorescent beads. The position measurements will be used to calculate shear stress auto-correlations. So far, however, the sensitivity of the measurements has proven to not be high enough to capture the predicted fluctuations. Future work in this direction will aim at increasing the softness of the microgel shells by reducing its degree of crosslinking to improve the sensitivity of the probe particles.

\begin{acknowledgements}
The authors thank Annette Zippelius and Jörg Baschnagel for helpful discussions.
MF thanks Peter Daivis for clarifying discussions and him and his colleagues from RMIT, Melbourne, for their hospitality  during a stay when this paper was written.
The work was supported  by the Deutsche Forschungsgemeinschaft (DFG) via SFB 1432 project C07. 
\end{acknowledgements}

\appendix
\section{Explicit expressions of the basis tensors}
\subsection{Three dimensions}\label{AppA3}
The basis orthogonal tesseral basis tensors $\tensor{\mathrm{Q}}^{(i)}$ are linear combinations of the following tensors $\tensor{\mathrm{C}}^{(i)}$
\begin{subequations}
\begin{gather}
   \tensor{\mathrm{C}}_1 =  \tensor{\delta} \; \tensor{\delta}\\
    \tensor{\mathrm{C}}_2 = \delta_{\alpha\delta}  \delta_{\gamma \beta} + \delta_{\alpha\gamma}\delta_{\beta\delta}    \\
    \tensor{\mathrm{C}}_3 = \tensor{\delta} \, \vec{\hat{r}}\,\vec{\hat{r}} + \vec{\hat{r}} \, \vec{\hat{r}} \,  \tensor{\delta}   \\
    \tensor{\mathrm{C}}_4   = \delta_{\alpha\delta}\, \hat{r}_\beta  \hat{r}_\gamma  + \delta_{\alpha\gamma}\hat{r}_\beta  \hat{r}_\delta + \delta_{\beta \delta}\hat{r}_\alpha\hat{r}_\gamma+ \delta_{\beta \gamma} {\hat{r}}_\alpha {\hat{r}}_\delta \\  
    \tensor{\mathrm{C}}_5 = \vec{\hat{r}}\,\uvec{r}\,\vec{\hat{r}}\,\vec{\hat{r}}
\end{gather}
\label{eq:ManuelBasis_3D}
\end{subequations}

connected by

\begin{gather}
    \begin{pmatrix}
        \tensor{\mathrm{Q}}_1 \\
        \tensor{\mathrm{Q}}_2 \\
        \tensor{\mathrm{Q}}_3 \\
        \tensor{\mathrm{Q}}_4 \\
        \tensor{\mathrm{Q}}_5  \\
    \end{pmatrix} = 
  \begin{pmatrix}
        \frac{1}{3} & 0 & 0 & 0 & 0 \\
        \frac{1}{3} & 0 & -\frac{1}{2} & 0 & 0 \\
        \frac{1}{6} & 0 & -\frac{1}{2} & 0 & \frac{3}{2} \\
         0 & 0 & 0 & \frac{1}{2\sqrt{2}} & -\sqrt{2} \\
         -\frac{1}{2\sqrt{2}} & \frac{1}{2\sqrt{2}} & \frac{1}{2\sqrt{2}} & -\frac{1}{2\sqrt{2}} & \frac{1}{2\sqrt{2}}
    \end{pmatrix}
    \begin{pmatrix}
        \tensor{\mathrm{C}}_1 \\
        \tensor{\mathrm{C}}_2 \\
        \tensor{\mathrm{C}}_3 \\
        \tensor{\mathrm{C}}_4 \\
        \tensor{\mathrm{C}}_5  \\
    \end{pmatrix} .  \label{eq:BasisLemaitre_3D}
\end{gather}
Orthogonality yields $\tensor{\mathrm{Q}}_i :\tensor{\mathrm{Q}}_j=\delta_{ij}$ simplifying the access to the isotropic functions by simple projections $q^{(i)}_{r,t} = \tensor{\sAC}(\vec{r},t):\tensor{\mathrm{Q}}_i(\hat{r})$, which as well holds in reciprocal space.

\subsection{Two dimensions}\label{AppA2}
In two dimensions some tensors in Eqs.\,\eqref{eq:ManuelBasis_3D} become dependent by the relation 
\begin{gather}
    \tensor{\mathrm{C}}_2 - \tensor{\mathrm{C}}_4 - 2\tensor{\mathrm{C}}_1 + 2 \tensor{\mathrm{C}}_3=0,
\end{gather}
making it possible to use only four of the tensors above in two dimensions.
The basis chosen in Ref.~\cite{Lemaitre_2D} reads 
\begin{gather}
    \begin{pmatrix}
        {\tensor{\mathrm{Q}} }_1 \\
        {\tensor{\mathrm{Q}} }_2 \\
        {\tensor{\mathrm{Q}}}_3 \\
        {\tensor{\mathrm{Q}}}_4 
    \end{pmatrix} = 
\begin{pmatrix}
         1/2 & 0 & 0 & 0 \\
         1/\sqrt{2} & 0 & -1/\sqrt{2} & 0 \\
         1/2 & 0 & -1 & 2 \\
         -1 & \frac{1}{2} & 1 & -2
    \end{pmatrix}
    \begin{pmatrix}
        \tensor{\mathrm{C}}_1 \\
        \tensor{\mathrm{C}}_2 \\
        \tensor{\mathrm{C}}_3 \\
        \tensor{\mathrm{C}}_5 
    \end{pmatrix}  \label{eq:BasisLemaitre_2D},
\end{gather}
with the two dimensional versions $\tensor{\mathrm{C}}_i$ of Eqs.\,\eqref{eq:ManuelBasis_3D}.

\section{Determination of sound velocities}
\label{AppB}
\begin{figure}[ht]
    \centering
    \includegraphics{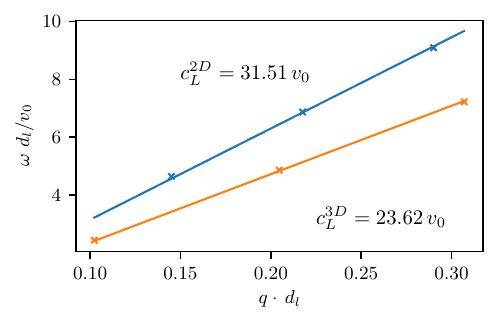}
    \caption{Newtonian system in two dimensions, dispersion relation. A linear fit gives the sound velocity $c_L$. Via $c_L=\sqrt{\frac{K_{pl}}{n}}$ one can determine the compression modulus or vice versa.}
    \label{fig:dispersionRelation_2D}
\end{figure}

To determine the sound velocity it was most convenient to measure the dispersion relation of the sound waves for a number of wave vectors. Therefore longitudinal velocity correlations were measured for three different wave-vectors. The frequency was determined by a damped harmonic oscillator fit, which then yields the dispersion relation. The sound velocity can then be read off the gradient of a linear fit. Figure \ref{fig:dispersionRelation_2D} shows the dispersion relation and the corresponding linear fit for two (blue) and three (orange) dimensions. The so determined sound velocities are noted in the figure.

\clearpage

\bibliography{apssamp}

\end{document}